# Interference and wave-particle duality of single photons

Shan-Liang Liu

Shandong Key Laboratory of Optical Communication Science and Technology, School of Physical Science and Information Engineering, Liaocheng University, Shandong 252059, China

The wave-particle duality has been said to contain the entire mystery of quantum mechanics. Many delayed-choice experiments have been performed to further understand the wave-particle duality. Here, we reveal some flaws in the known interference theories by comparing the theoretical predictions with the experimental facts and show that the presence of media is a necessary condition for interference of single photons, a photon interferes with other photon via microscopic particles in the interface of media, and the wave-like property of a photon originates from interaction between the photon and media; the particle property is the inherent characteristic of a photon, and the wave-like behavior of a photon always accompanies the particle behavior; the coherent time of single photons cannot be longer than the lifetime of the polarized states of the microscopic particles. Interference of single photons is interpreted properly, and the conundrum of the wave-particle duality is well solved.

While Huygens explained optical phenomena by a theory of waves, Newton put forward a corpuscular description where light consists of a stream of fast particles. However, Young's double-slit experiment demonstrated wave-like property of light at the beginning of the nineteenth century. Einstein revisited the theory of light acting as a particle to resolve conflicts between the wave theory of light and certain experimental results such as the photoelectric effect in 1905. Taylor used a sewing needle to split a beam of faint light from an incandescent source into two paths, on average only one photon at a time inside the apparatus, and observed interference fringes in 1909 [1]. This indicates that a photon has both wave-like and particle-like properties, i.e. wave-particle duality which has been said to contain the entire mystery of quantum mechanics [2,3]. Since the thermal light source do not generate photons one by one, Taylor's experiment can be described by the classical electromagnetic theory. Grangier and his coworkers performed the quantum interference of truly individual photons using a Mach-Zender interferometer (MZI) in 1986 [4].

A photon is sent toward MZI via the first 50:50 beam splitter (BS1), as shown in Fig. 1. The phase shift ϕ in the interferometer can be adjusted. When the phase shift between the two arms is varied, interference appears as a modulation of the detection probabilities at exits c and d of the second 50:50 beam splitter (BS2), respectively. This exhibits the wave-like property of a photon. If BS2 is removed, the photon has traveled only either path a or path b but not both after it passes the BS1, which fully exhibits the particle property of a photon and is incompatible with the classical electromagnetic theory [4,5], and no interference occurs. The photon antibunching characteristic originates from the particle property and has been become the crucial specification of a single-photon source [6,7]. So, a photon is impossible to interfere with itself, and how to interpret interference of single photons is still a puzzling question. Wheeler proposed the scheme of delayed-choice experiment where the decision to maintain or remove BS2 is delayed until after the photon passes BS1 to avoid the question whether a photon decides in advance to behave as a particle or a wave [8], as shown in Fig.1. Since then, many delayed-choice experiments have been performed [9]. Some experiments [10-12], in which whether BS2 is present or absent is decided after the photon passes BS1, were considered to support Bohr's complementary principle which states that a photon may behave either as a particle or a wave, depending on the measurement setup, but the two aspects appear to be incompatible and are never observed simultaneously [13]. The delayed-choice experiments with a quantum detecting setup show that a photon behaves both as a particle and as a wave simultaneously [14-16]. Recently, the wave and particle behaviors of a photon have been observed simultaneously in the delayed-choice experiment where BS2 can be either present or absent [13]. Therefore, how to understand the wave-particle duality of a photon is still a conundrum in physics. Here, we reveal some flaws in the known interference theories by comparing the theoretical predictions with the experimental facts and show that the presence of media is a necessary condition for interference of single photons, a photon interferes with other photon via the media, and the wave-like property of a photon originates from interaction between the photon and media; the particle property is the inherent characteristic of a photon.

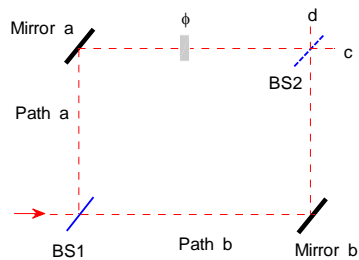

Fig. 1. Schematics of Wheeler's delayed-choice gedanken experiment.

Whether a photon is reflected into path a or transmitted into path b is random after it enters BS1 and is impossible to be predicted with certainty. The reflection and transmission probabilities depend on the boundary conditions and are derived by the classical electromagnetic theory. The wave-like property of a photon originates from interaction between the photon and the interface of BS1 and accompanies the particle property that the photon can be only in either path a or path b but not both. If BS1 is removed, the wave-like behavior of a photon never occurs, but the particle property still remains, and the photon reaches the mirror b along the path b with certainty. This means that the particle property is the inherent characteristic of a photon and is independent of whether media is present or absent. After a photon goes through BS1, the single-photon state can be described by the state vector

$$|\psi\rangle = \frac{1}{\sqrt{2}}(|1\rangle_a + |1\rangle_b), \quad (1)$$

where $|1\rangle_a$ or $|1\rangle_b$, corresponding to $|1\rangle_A|1\rangle_B$ or $|0\rangle_A|1\rangle_B$ in Ref. [17], denotes the state of the photon in path a or path b. A photon in $|\psi\rangle$ is in both $|1\rangle_a$ and $|1\rangle_b$ simultaneously according to the popular interpretation, but it contradicts with the experimental fact and is invalid. In fact, $|\psi\rangle$ describes the state of a single-photon ensemble in which there are a great number of single photons that enter BS1 at different times and have two possible states $|1\rangle_a$ and $||1\rangle_b$. Each photon in the ensemble can be in either $|1\rangle_a$ or $||1\rangle_b$ at the same probability after it goes through BS1, but it is impossible to occur in both the states as a wave simultaneously. The single-photon ensemble is mistaken as a photon in the popular interpretation. It is the misunderstanding that leads to the contradiction between the wave-like and particle-like properties. On the other hand, the media seems to play no role in the interference of single photons according to the known theories of interference [2,18], and the interference of single photons would occur whether BS2 is present or absent. However, this contradicts with the known experimental fact that the interference never occurs If BS2 is removed. It follows that the presence of BS2 is a necessary condition for the interference of single photons.

The electric fields in a linearly-polarized photon are in the same direction [19]. After a linearly-polarized photon with wave vector $k_a$ and the electric field $E_a$ from the pinhole a goes through a molecule at some point $r$ on the screen in Young's experiment, the molecule is polarized and has the electric dipole moment $p_a$, as shown in Fig. 2A. When a linearly-polarized photon with the wave vector $k_b$ and the electric field $E_b$ from the pinhole b meets with the polarized molecule, the electric dipole is subjected to a torque $p_a \times E_b$ and deflects toward the electric field $E_b$, and the latter deflects toward $p_a$ due to the opposite torque. After the photon goes through the polarized molecule, $p_a$ transforms into $p_b$, $E_b$ turns into $E_t$, $k_b$ becomes $k_t$, and the angle between the wave vectors $k_t$ and $k_a$ becomes larger than that between $k_b$ and $k_a$, as shown in Fig. 2B. Similarly, when a photon from the pinhole a meets with the polarized molecule with $p_b$, as shown in Fig. 2C, the electric dipole is subjected to a torque $p_b \times E_a$ and deflects toward $E_a$, and the latter deflects toward $p_b$ due to the opposite torque. After the photon goes through the polarized molecule, $p_b$ transforms into $p_a$, $E_a$ turns into $E_t$, $k_a$ becomes $k_t$, and the angle between $k_t$ and $k_b$ becomes larger than that between $k_a$ and $k_b$, as shown in Fig. 2D. It indicates that if the electric fields of two photons from the different pinholes are in the nearly opposite directions, the photons deviate from the original motion directions so that the angle between the wave vectors becomes larger after interaction with the polarized molecule. This corresponds to the destructive interference where the photons are deflected to the other places instead of being destroyed. Similarly, if the electric fields of the photons from the different pinholes are in nearly same direction, the photons slightly deviate from the original motion direction so that the angle between the wave vectors of the photons from the different pinholes becomes smaller after the interaction. This corresponds to constructive interference where there are the photons from the other positions. If the screen is removed, the polarization state and interference never occur.

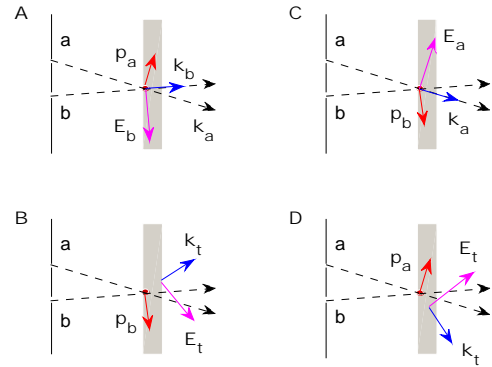

Fig. 2. Schematics of destructive interference in Young's experiment. A: the electric field $E_b$ and wave vector $k_b$ of the photon from pinhole b, and the dipole moment $p_a$ induced by the photon from pinhole a. B: the dipole moment $p_b$ after $p_a$ interacts with the photon from pinhole b, and the electric field $E_t$ and wave vector $k_t$ of the photon. C: the electric field $E_a$ and wave vector $k_a$ of the photon from pinhole a, D: the dipole moment $p_a$ after $p_b$ interacts with the photon from pinhole a, and the electric field $E_t$ and wave vector $k_t$ of the photon.

It is seen from the above analyses that the presence of screen is a necessary condition for interference of single photons, a photon interferes with other photon via the molecule in the surface of screen instead of itself, and the wave-like behavior of a photon always accompanies the particle behavior that a photon is only from either pinhole a or pinhole b but not both; it is not necessary for interference of single photons that the electric fields or wave functions of photons from different pinholes overlap in time domain.

A molecule makes a transition from the initial state $|i\rangle$ to the final state $|f\rangle$ due to interaction with a photon. The final state $|f\rangle$ only depends on the electric field of the photon and is independent of the initial state. The molecule polarization induced by a linearly-polarized photon from the pinhole a can be expressed as

$$E_a^{(\pm)}|i\rangle = e^{\pm i\varphi_a}|a\rangle, \quad (2)$$

where $|a\rangle$ is the normalized state vector of the polarized molecule with $\mathbf{p_a}$, and $E_a^{(\pm)}$ denotes the normalized operator of complex electric field of the photon and $\varphi_a$ is the phase [19]. Similarly, the molecule polarization induced by a linearly-polarized photon from the pinhole b can be expressed as

$$E_b^{(\pm)}|i\rangle = e^{\pm i\varphi_b}|b\rangle, \quad (3)$$

where $|b\rangle$ is the normalized state vector of the polarized molecule with $\mathbf{p_b}$, and $E_b^{(\pm)}$ denotes the normalized operators of complex electric field of the photon and $\varphi_b$ is the phase. In Young's experiment, the polarized molecule can be described by the normalized state vector

$$|i\rangle = \frac{1}{\sqrt{2}}(|a\rangle + |b\rangle), \quad (4)$$

and the normalized operator of complex electric field can be written as

$$E^{(+)} = \frac{1}{\sqrt{2}}(E_a^{(+)} + E_b^{(+)}). \quad (5)$$

Since both $|a\rangle$ and $|b\rangle$ never occur for a molecule simultaneously and are orthogonal each other. The normalized average intensity at the position **r** is given by

$$\langle i||E^{(+)}|^2|i\rangle = (1 + \cos\varphi_{ab})/2, \quad (6)$$

where $|E^{(+)}|^2 = E^{(-)}E^{(+)}$, and $\varphi_{ab} = \varphi_b - \varphi_a$ is the phase difference between the two photons which successively arrive at the position **r**. If $\mathbf{E_b}$ is nearly anti-parallel to $\mathbf{E_a}$, as shown in Fig. 2B, $\cos\varphi_{ab} \approx -1$, and the interference is full destructive. If $\mathbf{E_b}$ is nearly parallel to $\mathbf{E_a}$, $\cos\varphi_{ab} \approx 1$, and the interference is full constructive. The phase difference $\varphi_{ab}$ varies with the position **r**, and the interference pattern emerges. When one of two pinholes is blocked, no interference could occur if the photons arriving at any point on the screen were indistinguishable. Since the wave vectors of the photons from the different positions of the same pinhole are different in practice, the interference pattern still occurs, which is known as diffraction. Like interference, the presence of media is a necessary condition for diffraction.

The polarization state induced by the photon from the pinhole a at $t=t_a$ remains when the following photon from the pinhole b arrives at $t=t_b$ as long as $t_{ab}=t_b-t_a<t_p$ where $t_p$ denotes the lifetime of the polarization state, and interference can occur. The photon-induced polarization disappears if $t_{ab}>t_p$ and interference cannot occur. So, $t_p$ can be considered as the maximum of the coherent time and depends on the microscopic structure and temperature in the surface of screen.

Let $l_{ab}$ denote the path difference of two arms in MZI, the phase difference of the two photons at BS2 can be written as $\varphi_{ab}=kl_{ab}-\omega\Delta t$ where $\Delta t$ is the time interval of two photons which successively arrive at BS1. Whether a photon goes through MZI via path a or path b is known by no means, but the probability that the photon comes out from the exit c or d must $(1+\cos\varphi_{ab})/2$ or $(1-\cos\varphi_{ab})/2$. The wave-like behavior of a photon originates from interaction between the photon and the microscopic particles in the interface of BS2 and always accompanies the particle behavior that the photon can come out only from either exit c or exit d but not both. The wave-particle duality can be described by the state vector

$$|\psi'\rangle = \cos\frac{\varphi_{ab}}{2}|1\rangle_c + \sin\frac{\varphi_{ab}}{2}|1\rangle_d, \quad (7)$$

where $|1\rangle_c$ or $|1\rangle_d$ is the state vector of the photon from the exit c or d. If single photons are from an ideal single-photon source and the decoherence is negligible, $\Delta t$ is a constant, $\varphi_{ab}$ only varies with $l_{ab}$, and whether a photon comes out of the exit c to d can be controlled only by setting of the path difference $l_{ab}$. This is very significant for both quantum communication and quantum computation.

In conclusion, the presence of media is a necessary condition for interference of single photons, a photon interferes with other photon via microscopic particles in the interface of media, and the wave-like property of a photon originates from interaction between the photon and media. The particle property is the inherent characteristic of a photon, and the wave-like behavior of a photon always accompanies the particle behavior; the coherent time of single photons cannot be longer than the lifetime of the polarized states of the microscopic particles which depends on the microscopic structure and temperature of the interface. The interference of single photons is properly interpreted, and the conundrum of the wave-particle duality is well solved.